\documentclass[useAMS,usenatbib,usegraphicx]{mn2e}

\usepackage{amssymb}
\usepackage{times}

\def\apj{ApJ}%
\def\apjl{ApJ}%
\def\aap{A\&A}%
\def\prd{Phys.~Rev.~D}%
\def\pasj{PASJ}%
\def\physrep{Phys.~Rep.}%
\newcommand{\nue}{\ensuremath{\nu_{\rm e}}}
\newcommand{\nux}{\ensuremath{\nu_{\rm x}}}
\newcommand{\nuebar}{\ensuremath{\bar{\nu}_{\rm e}}}

\newcommand{\mdot}{\dot{M}}
\newcommand{\lnu}{L_{\nu}}
\newcommand{\lcore}{L_{\nue,\,{\rm core}}}
\newcommand{\msun}{\ensuremath{M_\odot}} 
\newcommand{\beq}{\begin{equation}}
\newcommand{\eeq}{\end{equation}}
\newcommand{\lcrit}{\lcore^{\rm crit}}
\newcommand{\vesc}{v_{\rm esc}}
\newcommand{\rnu}{r_\nu}
\newcommand{\rs}{\ensuremath{r_{\rm S}}}
\newcommand{\rscrit}{\ensuremath{\rs^{\rm crit}}}
\newcommand{\rsync}{\ensuremath{r_{\rm sync}}}
\newcommand{\rend}{\ensuremath{r_{\rm end}}}
\newcommand{\ye}{Y_{\rm e}}
\newcommand{\intd}{{\rm d}}
\newcommand{\invs}{{\rm s^{-1}}}
\newcommand{\erg}{{\rm ergs}}

\newcommand{\ergsec}{\ensuremath{\erg\ {\rm s}^{-1}}}
\newcommand{\epsnue}{\ensuremath{\varepsilon_{\nue}}}
\newcommand{\epsnuebar}{\ensuremath{\varepsilon_{\nuebar}}}
\newcommand{\epsnux}{\ensuremath{\varepsilon_{\nu_{\rm x}}}}
\newcommand{\phitot}{\ensuremath{\phi_{\nu,\bar{\nu}}}}
\newcommand{\fred}{\ensuremath{f_{\rm red}}}
\newcommand{\cno}{C$\nu$O}
\newcommand{\qdot}{\dot{q}}
\newcommand{\elec}{\ensuremath{e^-}}
\newcommand{\posi}{\ensuremath{e^+}}
\newcommand{\vff}{v_{\rm ff}}
\newcommand{\gf}{\ensuremath{G_{\rm F}}}
\defcitealias{pejcha12}{Paper I}

\title[Neutrino Oscillations and Supernovae]{Effect of Collective Neutrino Oscillations on the Neutrino Mechanism of Core-Collapse Supernovae}

\begin{document}
\date{Accepted ... Received ...}

\author[Pejcha, Dasgupta, \& Thompson]{\parbox[t]{\textwidth}{Ond\v{r}ej Pejcha$^1$, Basudeb Dasgupta$^2$ and Todd A. Thompson$^{1,2,3}$}
\vspace*{6pt}\\
$^1$Department of Astronomy, The Ohio State University, 140 West 18th Avenue, Columbus, OH 43210, USA \\
$^2$Center for Cosmology and Astroparticle Physics, The Ohio State University, 191 West Woodruff Avenue, Columbus, OH 43210, USA \\
$^3$Alfred P. Sloan Fellow \\ 
  email: pejcha@astronomy.ohio-state.edu}

\maketitle

\begin{abstract}
In the seconds after collapse of a massive star, the newborn proto-neutron star (PNS) radiates neutrinos of all flavors. The absorption of electron-type neutrinos below the radius of the stalled shockwave may drive explosions (the ``neutrino mechanism''). Because the heating rate is proportional to the square of neutrino energy, flavor conversion of $\mu$ and $\tau$ neutrinos to electron-type neutrinos via collective neutrino oscillations (\cno) may in principle increase the heating rate and drive explosions. In order to assess the potential importance of \cno\ for the shock revival, we solve the steady-state boundary value problem of spherically-symmetric accretion between the PNS surface ($\rnu$) and the shock ($\rs$), including a scheme for flavor conversion via \cno. For a given $\rnu$, PNS mass ($M$), accretion rate ($\mdot$), and assumed values of the neutrino energies from the PNS, we calculate the critical neutrino luminosity above which accretion is impossible and explosion results. We show that \cno\ can decrease the critical luminosity by a factor of at most $\sim\! 1.5$, but only if the flavor conversion is fully completed inside $\rs$ and if there is no matter suppression. The magnitude of the effect depends on the model parameters ($M$, $\mdot$, and $\rnu$) through the shock radius and the physical scale for flavor conversion. We quantify these dependencies and find that \cno\ could lower the critical luminosity only for small $M$ and $\mdot$, and large $\rnu$. However, for these parameter values \cno\ are suppressed due to matter effects. By quantifying the importance of \cno\ and matter suppression at the critical neutrino luminosity for explosion, we show in agreement with previous studies that \cno\ are unlikely to affect the neutrino mechanism of core-collapse supernovae significantly.
\end{abstract}


\section{Introduction}
\label{section:introduction}

Core-collapse supernovae announce the deaths of massive stars. The explosion develops deep in the optically thick stellar core and the dynamics of the neutron star in formation reveals itself only through the emission of neutrinos, which are scarcely detected \citep[e.g.][]{yuksel07}. Computer simulations  are thus the primary means to probe dynamics of the supernova explosions. Unfortunately, the non-rotating progenitors that explode have initial masses $\lesssim 15\,\msun$ \citep{rampp00,bruenn01,liebendorfer01,mezzacappa01,thompson03,kitaura06,janka08,fischer10,muller12}. In particular, it proves difficult to revive the outward movement of the shockwave, which forms when the collapsing core reaches nuclear density, and which stops its progress due to neutrino emission losses and other effects \citep[e.g.][]{burrows85,bruenn89a,bruenn89b}. Accretion of matter through the stalled shock ensues and lasts for many dynamical times before explosion or eventual black hole formation.  While rotation and magnetic fields help the explosion \citep[e.g.][]{leblanc70,symbalisty84,akiyama03,thompson05,dessart08,suwa10}, it is not likely that the magneto-rotational explosions are responsible for the majority of observed supernovae.

The hot proto-neutron star (PNS) cools by emission of neutrinos of all flavors. Because $\nu_\mu$, $\bar{\nu}_\mu$, $\nu_\tau$, and $\bar{\nu}_\tau$ (hereafter collectively denoted as $\nux$) do not interact with the PNS matter through the charged-current interactions, they decouple from matter at smaller PNS radii and higher temperatures, and thus can have higher average energy than $\nue$ and $\nuebar$. A fraction of the $\nue$ and $\nuebar$ are absorbed below the accretion shock and the associated energy deposition rate per unit mass is approximately proportional to $\lcore (\epsnue^2 + \epsnuebar^2)$, where $\lcore$ is the electron neutrino energy luminosity of the PNS, and $\epsnue$ and $\epsnuebar$ are the electron neutrino root-mean-square energies. This energy deposition plays a significant role in the dynamics of the supernova and perhaps in the revival of the shockwave \citep[e.g.][]{colgate66,bethewilson85}. Specifically, the ``neutrino mechanism'', as formulated by \citet{bg93}, states that the steady-state accretion through the shock turns into an explosion when $\lcore$ exceeds a critical value, $\lcrit$. In \citet{pejcha12} (hereafter \citetalias{pejcha12}) we showed using steady-state calculations that $\lcrit$ is equivalent to reaching $\max\,(c_S^2/\vesc^2) \simeq 0.19$ in the accretion flow, where $c_S$ is the sound speed and $\vesc$ is the local escape velocity. This ``antesonic'' condition is a manifestation of the inability of the flow to satisfy both the shock jump conditions and the Euler equations for the accretion flow simultaneously \citep{yamasaki05,fernandez12}. We also determined the dependence of $\lcrit$ on the key parameters of the problem, including the energies of the neutrinos over a wide range of parameter values. Specifically, and most importantly for this paper, we found that $\lcrit$ is proportional to the inverse square of the $\nue$ and $\nuebar$ energies, as expected from the heating rate. There are a number of time-dependent multi-dimensional effects that might modify the transition from accretion to explosion. For example, accretion luminosity from cooling of the accretion flow is an important contribution to $\lcrit$ \citepalias{pejcha12} and accretion simultaneously powering an asymmetric explosion is possible only in 2D and 3D \citep[e.g.][]{burrows06,marek09,suwa10}. Furthermore, close to the critical condition for explosion the shock surface often exhibits oscillations that feed back on the neutrino emission \citep[e.g.][]{murphy08,marek09,nordhaus10,hanke11} potentially modifying $\lcrit$. At least in 1D, these oscillations seem to occur only very close to the steady-state value of $\lcrit$ \citep{fernandez12}. The steady-state calculation is thus useful way to estimate $\lcrit$ and to examine effects of modified physics on the critical condition for supernova explosion.

Within the parameterization of the neutrino mechanism of \citet{bg93}, the failure of supernova simulations implies, by definition, that the neutrino luminosities in the models never reach $\lcrit$. For successful explosions, either (i) $\lcrit$ needs to be decreased or (ii) $\lcore$ increased. As an example of the former, multi-dimensional effects like convection and SASI decrease $\lcrit$ \citep{yamasaki05,yamasaki06,murphy08,nordhaus10,hanke11,takiwaki12} by making the heating more efficient \citep[e.g.][]{herant94,bhf95,janka96,fryer04,buras06a}. As an example of the latter, $\lcore$ can be enhanced by convection inside the PNS \citep[e.g.][]{wilson88,bruenn96,keil96}. Another option is that cooling becomes less efficient in $2$ and $3$ spatial dimensions causing a decrease of $\lcrit$ \citepalias{pejcha12}. 

Most of the heating below the shock occurs due to absorption of $\nue$ and $\nuebar$ on neutrons and protons, while the $\nux$ escape without much interaction. However, due to the high density of neutrinos in this region, self-interaction between neutrinos becomes important and can lead to a range of phenomena called ``collective neutrino oscillations'' \citep[e.g.][]{pantaleone92,duan06,duan10}. In particular, there is a possibility of an instability \citep{dasgupta09} that exchanges part of the $\nux$ spectra with the $\nue$ and $\nuebar$ spectra. If the luminosities and energies of $\nue$, $\nuebar$, and $\nux$ are right, this can produce significantly more heating below the shock than calculations neglecting neutrino oscillations, i.e., effective $\lcore$ is increased by \cno. Specifically, a strong effect on heating can be expected if luminosities are similar and $\nux$ have significantly higher energies than $\nue$ and $\nuebar$. The exact values of these quantities and their mutual ratios are model-dependent \citep[e.g.][]{thompson03,marek09,fischer10,fischer12,hudepohl10}. 

\citet{chakraborty11a,chakraborty11b}, \citet{dasgupta12}, \citet{suwa11}, and \citet{sarikas11} have investigated the role of \cno\ in the core-collapse simulations of several progenitor models. They found that there are a number of multi-angle effects, especially the effect of matter suppression, that can reduce or entirely eliminate \cno. We address the issue of increased neutrino heating due to \cno\ and matter suppression without reference to detailed supernova models and we evaluate them at $\lcrit$, which separates accretion from explosion. We treat the oscillation physics in a schematic way and the supernova physics using a steady state model developed in \citetalias{pejcha12}. Although this approach is less detailed than some recent papers, e.g., \citep[e.g.][]{chakraborty11a,chakraborty11b,dasgupta12,sarikas11}, it allows for a parametric study to determine the potential role of \cno\ in shock reheating, and its dependence on the progenitor mass, radius, and accretion rate for a very broad range of parameters and without being tied to any particular progenitor model or simulation setup.

The remainder of our paper is organized as follows. In Section~\ref{sec:method}, we describe our steady state model for the accretion flow based on \citetalias{pejcha12}, and a scheme for collective neutrino oscillations based on \citet{dasgupta12}. 
We present our results in Section~\ref{sec:results}. We quantify the changes to $\lcrit$ and the shock radii, and compare the magnitude of the effect of \cno\ to other known pieces of physics. We also estimate the importance of multi-angle effects showing that they suppress \cno\ in the region of parameter space where they might otherwise be strong.
In Section~\ref{sec:disc}, we conclude with a discussion and review of our results.

\section{Method}
\label{sec:method}

In this Section we first describe the hydrodynamic equations that we shall solve, their boundary conditions, and the input neutrino physics (Section~\ref{sec:hydro}). We describe our scheme of coupling the \cno\ effects to the hydrodynamical equations in Section~\ref{sec:cno}. Our combination of steady-state approach and simple treatment of \cno\ allows us to calculate $\lcrit$, quantify the maximum possible effect of \cno\ on $\lcrit$ as a function of boundary conditions, and set limits on the parameter space, which can then be probed with more realistic methods.

\subsection{Hydrodynamic Equations and Boundary Conditions}
\label{sec:hydro}

We use the code developed in \citetalias{pejcha12} to calculate the structure of the steady-state accretion flow between the neutrinosphere at radius $\rnu$ and the standoff accretion shock at $\rs$ assuming spherical symmetry by solving the time-independent Euler equations
\begin{eqnarray}
\frac{1}{\rho} \frac{\intd \rho}{\intd r} +\frac{1}{v} \frac{\intd v}{\intd r} + \frac{2}{r} &=& 0, \label{eq:mass_cons}\\
v\frac{\intd v}{\intd r} + \frac{1}{\rho}\frac{\intd P}{\intd r} &=& -\frac{GM}{r^2}, \label{eq:moment_equ}\\
\frac{\intd\mathcal{E}}{dr} - \frac{P}{\rho^2}\frac{\intd\rho}{\intd r} &=& \frac{\qdot}{v},\label{eq:energy_tot}
\end{eqnarray}
where $P$ is the gas pressure, $M$ is the mass within the radius of the neutrinosphere $\rnu$, $\mathcal{E}$ is the internal specific energy of the gas, and $\qdot$ is the net heating rate, a difference of heating and cooling. We solve for the electron fraction $\ye$ using
\beq
v\frac{\intd\ye}{\intd r} = l_{\nue n} + l_{\posi n} - (l_{\nue n} + l_{\posi n} + l_{\nuebar p} + l_{\elec p})\ye \label{eq:el_frac},
\eeq
where $l_{\nue n}$, $l_{\posi n}$, $l_{\nuebar p}$, and $l_{\elec p}$ are reaction rates involving neutrons and protons. These equations are coupled together through the equation of state (EOS), $P(\rho, T, \ye)$ and $\mathcal{E}(\rho, T, \ye)$, where $T$ is the gas temperature. Our EOS contains relativistic electrons and positrons, including chemical potentials, and nonrelativistic free protons and neutrons \citep{qian96}. We use prescriptions of heating, cooling, opacity and reaction rates for charged-current processes with neutrons and protons from \citet{schecketal06}. We do not include the accretion luminosity, i.e.\ the changes in neutrino luminosities as a function of radius due to emission or absorption of neutrinos, but our neutrino luminosities change as a function of radius due to the \cno. Here, we also assume that $\nux$ do not directly interact with matter, but instead they convert to $\nue$ and $\nuebar$ through \cno\ as described in Section~\ref{sec:cno}. While we explicitly calculate the degeneracy parameter of electrons and positrons, we assume that the degeneracy parameter of (anti)neutrinos is zero.

We need five boundary conditions to uniquely determine four functions: $\rho$, $v$, $T$ and $\ye$, and shock radius $\rs$. We first demand that the flow has a fixed mass accretion rate $\mdot = 4\pi r^2 \rho v$ by applying this constraint at the inner boundary. At the outer boundary we apply the shock jump conditions
\begin{eqnarray}
\rho v^2 + P &=& \rho^+ \vff^2,\label{eq:shock_jump_mom}\\
\frac{1}{2}v^2 + \mathcal{E} + \frac{P}{\rho} &=& \frac{1}{2} \vff^2,\label{eq:shock_jump_ene}
\end{eqnarray}
where $\vff = \sqrt{\Upsilon} v_{\rm esc}$ is the free fall velocity, and we choose $\Upsilon = 0.25$ in agreement with the analytically estimated mass accretion rates of \citet{woosley02} supernova progenitors. The quantity $\rho^+$ is the density just upstream of the shock and can be calculated from conservation of mass. We also assume that the matter entering the shock is composed of iron and hence $\ye = 26/56$ at the outer boundary. The last boundary condition comes from requiring that $\rnu$ is the neutrinosphere for electron neutrinos, which gives a constraint on the optical depth
\beq
\tau_{\nue} = \int_{\rnu}^{\rs} \kappa_{\nue} \rho\,\,\intd r = \frac{2}{3},
\label{eq:taudef}
\eeq
where $\kappa_{\nue}$ is the opacity to electron neutrinos, which is proportional to the mean square neutrino energy $\epsnue^2$.

The key parameters of the problem are the mass accretion rate $\mdot$ through the shock, the PNS mass $M$, its radius $\rnu$, and the core neutrino luminosity of each species $\lcore$. The shock radius $\rs$ is the eigenvalue of the problem and is determined self-consistently in the calculation\footnote{The only difference in the numerical setup in this paper with respect to \citetalias{pejcha12} is that here we assume that luminosities in each neutrino species are constant throughout the region of interest. In \citetalias{pejcha12} we implemented a simple gray neutrino transport, which allows for $\intd \lnu / \intd r \neq 0$. There is also a difference in notation in the sense that $L_{\nu,{\rm core}}$ in \citetalias{pejcha12} refers to the combined electron neutrino and antineutrino luminosity, while here we parameterize the luminosities separately, $L_{\nu,{\rm core}} = 2\lcore$.}. The critical luminosity $\lcrit$ is determined by increasing $\lcore$ with the remaining parameters fixed until no steady-state solution is possible. For $\lcore > \lcrit$, the solution most likely transitions to a neutrino-driven wind, which is identified with the supernova explosion \citep{burrows87,bg93,yamasaki05,fernandez12}. The highest value of $\lcore$ that yields a steady-state solution is identified with $\lcrit$. The value of $\lcrit$ is polished to the desired level of precision by using a bisection method. 

In \citetalias{pejcha12} we calculated $\lcrit$ and $\rs(\lcore)$ for a wide range of $\mdot$, $M$, and $\rnu$. We found that $\lcrit$ scales approximately as a power law\footnote{We argued in \citetalias{pejcha12} that the small upward curvature of $\lcrit(\mdot)$ is caused most likely by an exponential correction factor that arises due to nearly-hydrostatic structure of the accretion flow.} in $\mdot$, $M$, and $\rnu$,
\beq
\lcrit \propto \mdot^{0.723} M^{1.84} \rnu^{-1.61},
\label{eq:lcrit_empir}
\eeq
for $0.01\le \mdot \le 2\,\msun$\,s$^{-1}$, $1.2\le M\le 2.0\,\msun$, and $20\le \rnu \le 60$\,km. We also found that for the physical solutions, $\rs$ increases with increasing $\lcore$ \citep{yamasaki05} and that the values of $\rs$ at $\lcrit$, $\rscrit$, depend only on the actual value of $\lcrit$. Specifically, we found that 
\beq
\frac{\rscrit}{\rnu} \propto (\lcrit)^{-0.26}.
\label{eq:rscrit}
\eeq

\subsection{Treatment of \cno}
\label{sec:cno}

In this Section, we describe our implementation of \cno\ effects in our calculation. In order to evaluate the maximum potential effect on $\lcrit$, we employ a simple treatment of \cno\ that is relatively insensitive to details of, or approximations to, otherwise complicated calculations of \cno. Essentially, we build on the results of \citet{dasgupta12} who showed that this simple approach is a good approximation to more complicated calculations, and we extend their work to a larger parameter space $\mdot$, $M$, and $\rnu$, and evaluate \cno\ at $\lcrit$, the boundary between accretion and explosion.

To calculate the effective change of neutrino energies and luminosities as a function of radius we assume that a maximum possible flavor conversion occurs and we model this as a smooth non-oscillatory transition of the flavors. More specifically, in \cno\ the neutrinos convert to other flavors through interactions schematically written as 
\beq
\nue\nuebar \leftrightarrows \nux\bar{\nu}_{\rm x}.
\label{eq:nu_react}
\eeq
With maximum possible flavor conversion, the ``final'' neutrino number fluxes $\phi_{\nue}^{\rm f}$, $\phi_{\nuebar}^{\rm f}$, and $\phi_{\nux}^{\rm f}$ can be expressed using the ``initial'' neutrino number fluxes $\phi_{\nue}^{\rm i}$, $\phi_{\nuebar}^{\rm i}$, and $\phi_{\nux}^{\rm i}$ at the neutrino sphere. For example if $\phi_{\nue}^{\rm i} > \phi_{\nuebar}^{\rm i}$ (which is always true for the scenarios considered in this paper), then from Equation~(\ref{eq:nu_react}) follows that the maximum number flux of $\nue$ that can be converted to $\nux$ is equal to $\phi_{\nuebar}^{\rm i}$. Similarly, one quarter of the total flux in $\nux$, $\phi_{\nux}^{\rm i}$, is converted to $\nue$ and one quarter to $\nuebar$; the remaining half stays in $\nux$. After the conversion is fully done, the $\nue$ number flux is a sum of $\phi_{\nue}^{\rm i}-\phi_{\nuebar}^{\rm i}$ with the original energy $\epsnue$, which could not have been converted due to a lack of $\nuebar$, plus the number flux $\phi_{\nux}^{\rm i}$ of converted $\nux$ neutrinos with energy $\epsnux$. Similar logic gives the final number fluxes in all flavors:
\begin{eqnarray}
\phi_{\nue}^{\rm f} &=& \phi_{\nue}^{\rm i} - \phi_{\nuebar}^{\rm i} + \phi_{\nux}^{\rm i},\label{eq:phi_nuef}\\
\phi_{\nuebar}^{\rm f} &=& \phi_{\nux}^{\rm i},\\
4\phi_{\nux}^{\rm f} &=&  2\phi_{\nux}^{\rm i} + \phi_{\nuebar}^{\rm i} + \phi_{\nuebar}^{\rm i},\label{eq:phi_nuxf}
\end{eqnarray}
where in the equation for $\phi_{\nux}^{\rm f}$ one instance of $\phi_{\nuebar}^{\rm i}$ represents converted $\nue$ with the appropriate energy and the other one converted $\nuebar$ with their energy. The total neutrino number flux $\phitot = \phi_{\nue} + \phi_{\nuebar} + 4\phi_{\nux}$ is conserved, specifically $\phitot^{\rm i} = \phitot^{\rm f}$. The rms neutrino energies in each flavor after the conversion is finished, $\epsnue^{\rm f}$, $\epsnuebar^{\rm f}$, and $\epsnux^{\rm f}$, are calculated as rms of the initial energies weighted by the fluxes as given in Equations~(\ref{eq:phi_nuef})--(\ref{eq:phi_nuxf}), specifically for the electron-flavor neutrinos $(\epsnue^{\rm f})^2 =  \left[ (\phi_{\nue}^{\rm i} - \phi_{\nuebar}^{\rm i})(\epsnue^{\rm i})^2 + \phi_{\nux}^{\rm i}(\epsnux^{\rm i})^2\right]/\phi_{\nue}^{\rm f}$ and $\epsnuebar^{\rm f} = \epsnux^{\rm i}$.

For $\nue$, the number flux at the neutrinosphere $\phi_{\nue}^{\rm i}$ is related to the core energy luminosity $\lcore$ as
\beq
4\pi\rnu^2\phi_{\nue}^{\rm i} = \frac{\lcore}{\langle\epsnue^{\rm i}\rangle},
\label{eq:phi}
\eeq
where $\langle \epsnue^{\rm i} \rangle$ is the mean energy of $\nue$ at the neutrinosphere. Expressions similar to Equation~(\ref{eq:phi}) are valid for $\phi_{\nuebar}^{\rm i}$ and $\phi_{\nux}^{\rm i}$ with initial mean energies $\langle\epsnuebar^{\rm i}\rangle$ and $\langle \epsnux^{\rm i}\rangle$. In order to calculate the mean energies from the rms energies, that we use throughout the paper, we assume that the neutrino energy spectrum is Fermi-Dirac, which gives the necessary conversion factor. The differences in the number fluxes when compared to using the rms energies are $\sim 13\%$, but as we show below the dependence of physical scale of conversion on the number flux is rather weak. The neutrinosphere radius $\rnu$ of $\nue$ neutrinos is used to calculate the number fluxes at $\rnu$ for all neutrino flavors, because at $\rnu$ the $\nuebar$ and $\nux$ are already essentially free-streaming.

Motivated by more complete studies of the physical extent of the flavor conversion \citep[e.g.][]{dasgupta12}, and as a numerical expedient, we assume that the neutrino number fluxes and energies for each flavor vary smoothly as a function of radius between the initial and final states. For $\nue$,
\begin{eqnarray}
\phi_{\nue}(r) &=& \left[1-P(r)\right]\phi_{\nue}^{\rm i} + P(r)\phi_{\nue}^{\rm f},\\
\epsnue(r) &=& \left[1-P(r)\right]\epsnue^{\rm i} + P(r)\epsnue^{\rm f},
\end{eqnarray}
and similar equations are valid for $\nuebar$ and $\nux$. Neutrino luminosity $L_{\nue}(r)$ as a function of radius is calculated as a product of $\phi_{\nue}(r)$ and $\epsnue(r)$, which guarantees that $L_{\nue}(\rnu) = \lcore$. The ``survival probability'' $P(r)$ is modelled after the full solution to the \cno\ problem of \citet{dasgupta12} as
\beq
P(r) = \frac{1}{2}+\frac{1}{2}\tanh\left[ \frac{2r- (\rsync+\rend)}{\sigma(\rend-\rsync)}  \right],\label{eq:surv_prob}
\eeq
where we choose $\sigma \simeq 0.679$ to have $P(\rsync) = 0.05$ and $P(\rend) = 0.95$. We shall use  simple estimates for $\rsync$ and $\rend$ of flavor conversion derived from the analysis in \citet{hannestad06}, and summarized in \citet{dasgupta12}. This approach is the most optimistic possibility for oscillations. Our aim in this paper is to ascertain the \emph{maximum} effect that \cno\ may have on the supernova shock reheating, and therefore our constraints shall be conservative. Any new physics effects that reduce the effect of \cno\ will only strengthen the constraints we derive here.

The two parameters $\rsync$ and $\rend$ in Equation~(\ref{eq:surv_prob}) define the range of radii where \cno\ operates in a simplified treatment of nonlinear effects of neutrino-neutrino interactions \citep{hannestad06}. The flavor conversion occurs above a synchronization radius $\rsync$, which is defined as
\beq
\mu(\rsync) = 4\Omega \left( 1+\frac{\rnu^2}{4\rsync^2} \right).
\label{eq:rsync}
\eeq
The flavor conversion is more or less complete at radius $\rend$ defined as
\beq
\mu(\rend) = \Omega \frac{\mathcal{F}_-}{\mathcal{F}_+} \left( 1+\frac{\rnu^2}{4\rend^2}\right).
\label{eq:rend}
\eeq
Here, $\Omega$ depends on the neutrino oscillation frequency and neutrino energy spectra and as in \citet{dasgupta12}, we choose a typical value $\Omega =50$\,km$^{-1}$. The quantity $\mathcal{F}_-/\mathcal{F}_+$ is the ratio of the net lepton asymmetry in the system $\mathcal{F}_-$ to the neutrino flux available for oscillations $\mathcal{F}_+$, which are defined as
\begin{eqnarray}
\mathcal{F}_- &=& \frac{\phi_{\nue}^{\rm i}-\phi_{\nuebar}^{\rm i}}{\phitot}, \label{eq:fminus}\\
\mathcal{F}_+ &=& \frac{\phi_{\nue}^{\rm i} + \phi_{\nuebar}^{\rm i} - 2\phi_{\nux}^{\rm i}}{\phitot}. \label{eq:fplus}
\end{eqnarray}
The collective potential $\mu$ is defined as \citep{esteban07,dasgupta12}
\beq
\mu(r) = \sqrt{2}\gf \phitot \left(\frac{\rnu}{r}\right)^2 \left( \frac{\rnu^2}{2r^2-\rnu^2} \right),
\label{eq:mu}
\eeq
where $\gf$ is the Fermi coupling constant. Equations~(\ref{eq:rsync}) and (\ref{eq:rend}) are solved for $\rsync$ and $\rend$ given $\rnu$ and the neutrino number fluxes. We note here that the dependence of Equation~(\ref{eq:phi}) on $\rnu$ introduces an absolute scale to Equations~(\ref{eq:rsync}) and (\ref{eq:rend}), and therefore $\rsync$ and $\rend$ do not scale linearly with $\rnu$. Instead, in the limit of $\rsync/\rnu \gg 1$, the scaling is 
\beq
\frac{\rsync}{\rnu} \propto \lcore^{1/4}\rnu^{-1/2}
\label{eq:scaling}
\eeq
for fixed neutrino energies. The same scaling holds for $\rend$.

The collective neutrino oscillations can be suppressed due to a variety of effects. Out of them, the most relevant here is the matter suppression of the \cno\ \citep{esteban08}. This occurs when the Mikheyev-Smirnov-Wolfenstein (MSW) potential $\lambda$ becomes much greater than a critical value $\lambda_{\rm MA}$. The MSW potential is 
\beq
\lambda = \sqrt{2} \gf \left(n_{\rm e^-} - n_{\rm e^+}\right) = \sqrt{2} \gf \frac{\ye\rho}{m_{\rm n}},
\label{eq:lambda}
\eeq
where $n_{\rm e^-}$ and $n_{\rm e^+}$ are the number densities of electrons and positrons, respectively, and we assume that the matter is composed only of protons, neutrons, electrons, and positrons. The critical value $\lambda_{\rm MA}$ is defined as
\beq
\lambda_{\rm MA} = 2\sqrt{2} \gf \phitot \frac{\rnu^2}{r^2} \mathcal{F}_-,
\label{eq:lambda_MA}
\eeq
and it is essentially the collective potential weighted by the lepton asymmetry factor \citep{dasgupta12}.

We emphasize that in the calculation presented here, the neutrino energies and luminosities as a function of radius self-consistently enter not only in the heating, but also in the reaction rates for the calculation of $\ye$ (Eq.~[\ref{eq:el_frac}]) and in the boundary condition on optical depth (Eq.~[\ref{eq:taudef}]). However, in our approximation we do not include accretion luminosity and we cannot resolve any of the multi-dimensional effects, although we attempt to effectively capture them by modifying the cooling rates as will be discussed in Section~\ref{sec:cno_lcrit}.

\section{Results}
\label{sec:results}

In this Section, we illustrate the effect of \cno\ on the critical luminosity required for shock revival. First, we show our results with $(\epsnue^{\rm i}, \epsnuebar^{\rm i},  \epsnux^{\rm i}) = (13,15.5,20)$\,MeV \citep{thompson03}, and $\lcore = L_{\nuebar,{\rm core}} = L_{\nux,{\rm core}}$, as an optimistic scenario for \cno\ (Section~\ref{sec:cno_lcrit}). Then we study a more realistic scenario with $(\epsnue^{\rm i}, \epsnuebar^{\rm i},  \epsnux^{\rm i}) = (11,13,18)$~MeV, and $\lcore = L_{\nuebar,{\rm core}} = 2L_{\nux,{\rm core}}$, based on recent detailed simulations of the accretion phase \citep[e.g.][]{marek09,fischer10,fischer12}. We discuss our results primarily for the former case, but we find that the latter case differs only in the maximum strength of \cno\ and not in the parameter range. We assess \cno\ for different sets of parameters in Section~\ref{sec:comparison}. We also initially ignore the matter suppression throughout the discussion to illustrate the unattenuated magnitude of the effect, but we return to the matter suppression in Section~\ref{sec:multiangle} and show how it further constrains the parameter space for \cno.

\subsection{Effect of \cno\ on $\lcrit$}
\label{sec:cno_lcrit}

\begin{figure*}
\centering
\includegraphics[width=0.5\textwidth]{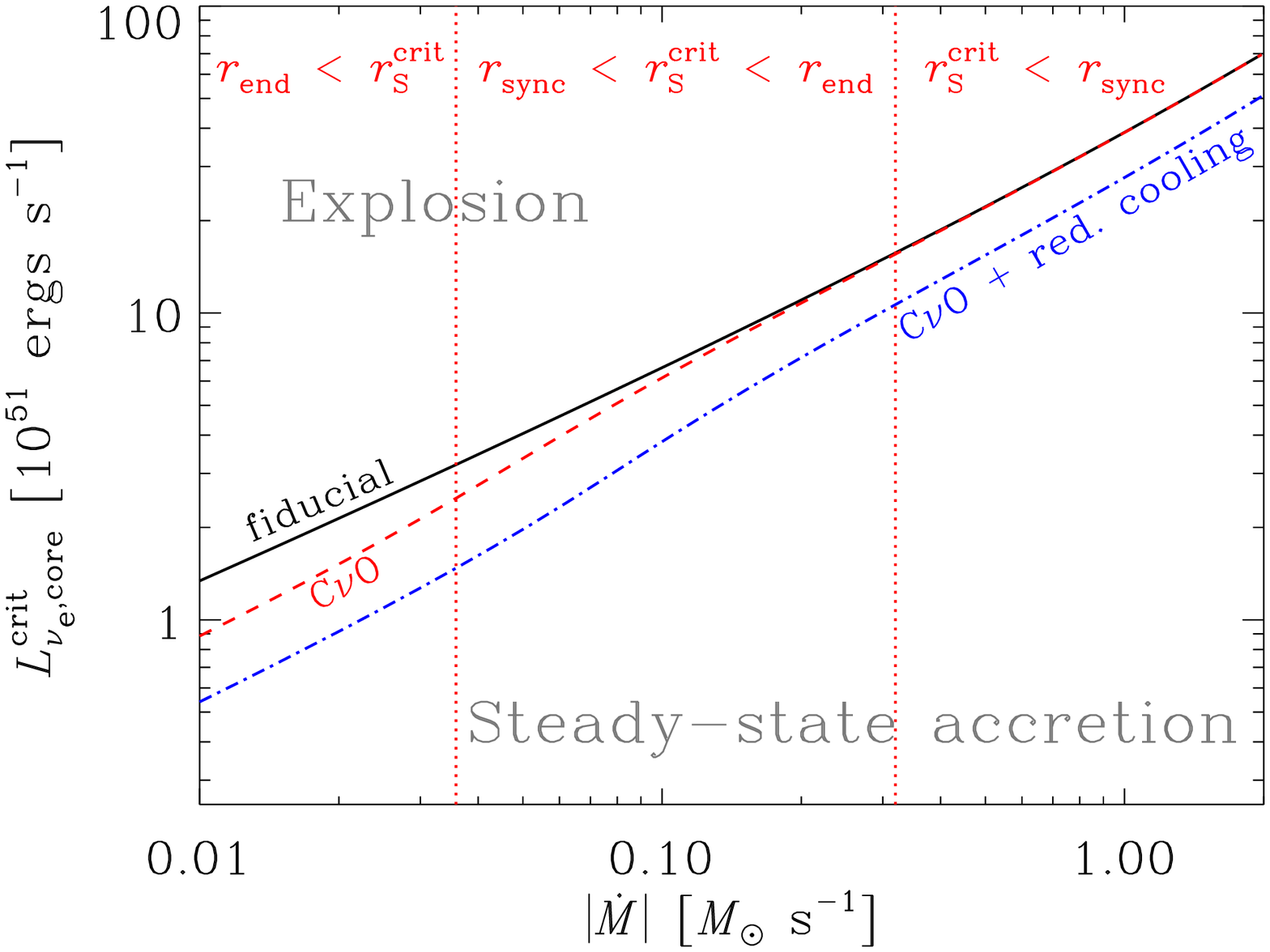}\hfill\includegraphics[width=0.5\textwidth]{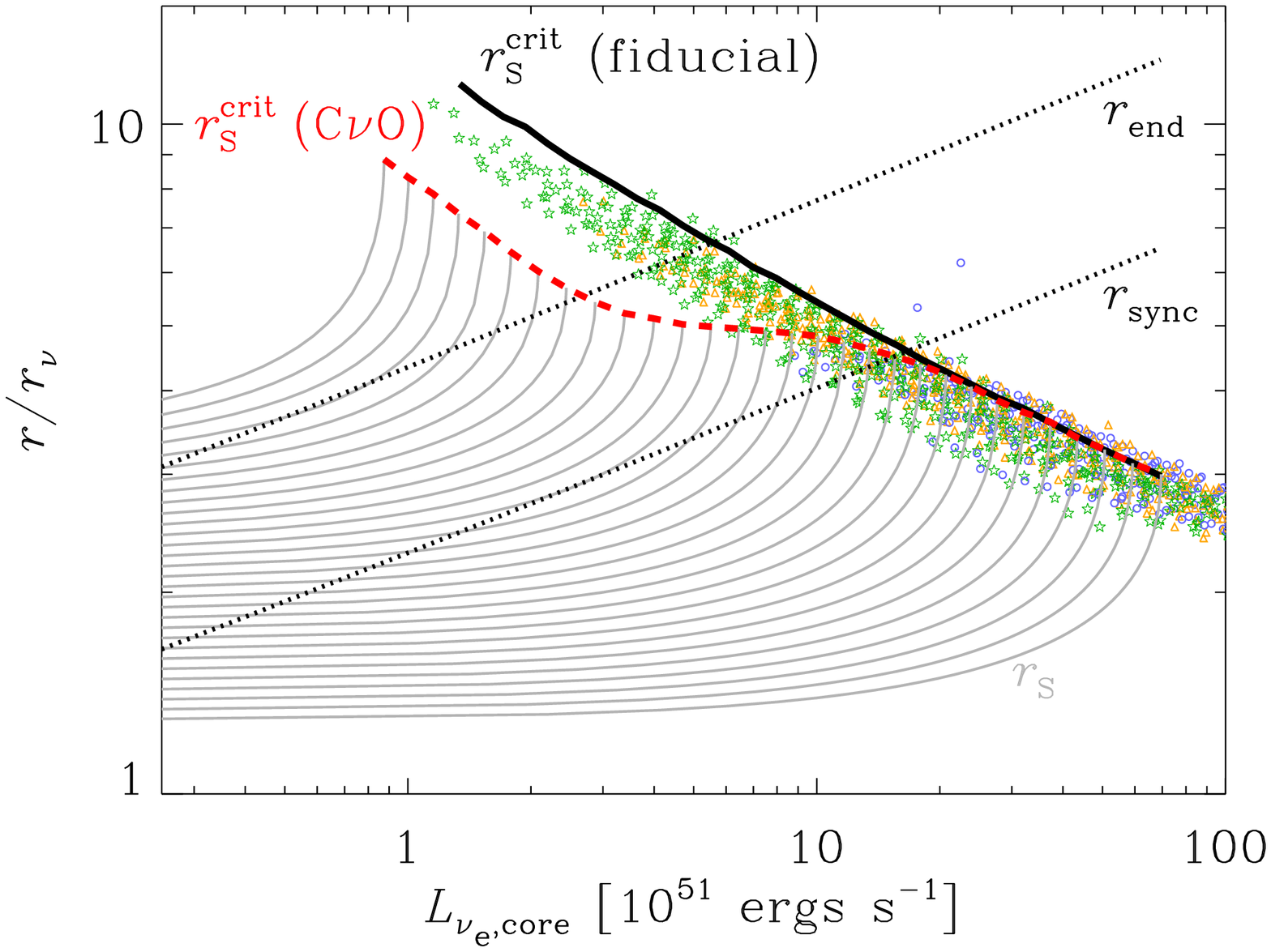}
\caption{{\em Left}: critical core neutrino luminosity $\lcrit$ as a function of $\mdot$ for $M=1.2\,\msun$ and $\rnu=60$\,km. The solid black line shows the fiducial calculation (no \cno), while the red dashed line is with \cno\ included. The vertical red dotted lines mark different regimes of the effect (see text and the right panel). The blue dash-dotted line includes \cno\ and has the neutrino cooling reduced by a factor of $2$ to approximate multi-dimensional effects. {\em Right}: effect of \cno\ on shock radii. Dots mark $\rscrit$ for calculations from \citetalias{pejcha12}, illustrating that $\rscrit$ depends primarily on $\lcrit$. Blue dots, orange triangles, and green stars are for $\rnu=20$, $40$, and $60$\,km, respectively. The black solid line shows $\rscrit$ for the fiducial calculation from this paper for the same $M$ and $\rnu$ as in the left panel. The grey lines show $\rs$ as a function of $\lcore$ for different $\mdot$ with \cno\ included. The grey lines terminate at $\rscrit$, which are connected with a dashed red line (compare with red dashed line, left panel). The black dotted lines show $\rsync$ and $\rend$. For $\rscrit < \rsync$, \cno\ has no effect; for $\rscrit > \rend$, the effect of the \cno\ is maximized.}
\label{fig:osc}
\end{figure*}

In the left panel of Figure~\ref{fig:osc}, we show $\lcrit$ for $(\epsnue^{\rm i}, \epsnuebar^{\rm i},  \epsnux^{\rm i}) = (13,15.5,20)$\,MeV and $\lcore = L_{\nuebar,{\rm core}} = L_{\nux,{\rm core}}$ including \cno\ as a function of $\mdot$ for $M=1.2\,\msun$ and $\rnu=60$\,km (red dashed line) along with the fiducial calculation without the effect of \cno\ (black solid line). The critical curve in the fiducial calculation is approximately a power law \citepalias{pejcha12}. We see that \cno\ lowers $\lcrit$ to $\sim\! 0.65$ times the fiducial value for $\mdot < 0.01\ \msun\ \invs$. For higher $\mdot$, the critical curve turns upward and for $\mdot \gtrsim 0.32\ \msun\ \invs$, it essentially coincides with the fiducial calculation meaning that \cno\ have little effect.

The behavior seen in the left panel of Figure~\ref{fig:osc} is non-trivial even in our simple setup, because $\rsync$ and $\rend$ are a function of the boundary conditions. It can be understood by analyzing the position of $\rs$ relative to $\rsync$ and $\rend$. We expect that \cno\ will have an effect on $\lcrit$ only if the conversion starts below the shock radius and the full effect will be obtained if the conversion is completed inside the standing shock. For the parameters we consider here, $\rsync$ is always above the neutrinosphere. Because $\rs$ increases with $\lcore$ (grey solid lines in Fig.~\ref{fig:osc}, right panel) and reaches a maximum $\rscrit$ at $\lcrit$, the effect of \cno\ is most prominent for $\lcore$ close to $\lcrit$.  The right panel of Figure~\ref{fig:osc} shows the effect of \cno\ on the shock radii. We see that $\rscrit$ in the fiducial calculation (thick black solid line) closely follows the results from \citetalias{pejcha12} (Eq.~[\ref{eq:rscrit}], dots, triangles and stars in Figure~\ref{fig:osc}) except that the calculations presented here have larger $\rscrit$, because we set $\intd \lnu / \intd r = 0$. The calculation with \cno\ (red dashed line) closely follows the fiducial results for high $\mdot$ and $\lcrit$, where $\rsync > \rscrit$. Here, the \cno\ effect is negligible and $\lcrit$ is very close to the fiducial value. When $\rscrit \simeq \rsync$, which occurs for $\lcrit \simeq 1.5\times 10^{52}$\,\ergsec\ ($\mdot \simeq 0.32\ \msun\ \invs$), the \cno\ effect starts to become important, and both $\lcrit$ and $\rs$ decrease relative to the fiducial calculation. For $\lcrit \lesssim 2.5 \times 10^{51}$\,\ergsec, which corresponds to $\mdot \lesssim 0.035\ \msun\ \invs$, $\rend < \rscrit$ and \cno\ affect the structure of the flow significantly and essentially saturate. $\lcrit$ with \cno\ is reduced to $\sim\! 0.65$ of the fiducial value for low $\mdot$.

\begin{figure*}
\centering
\includegraphics[width=0.6\textwidth]{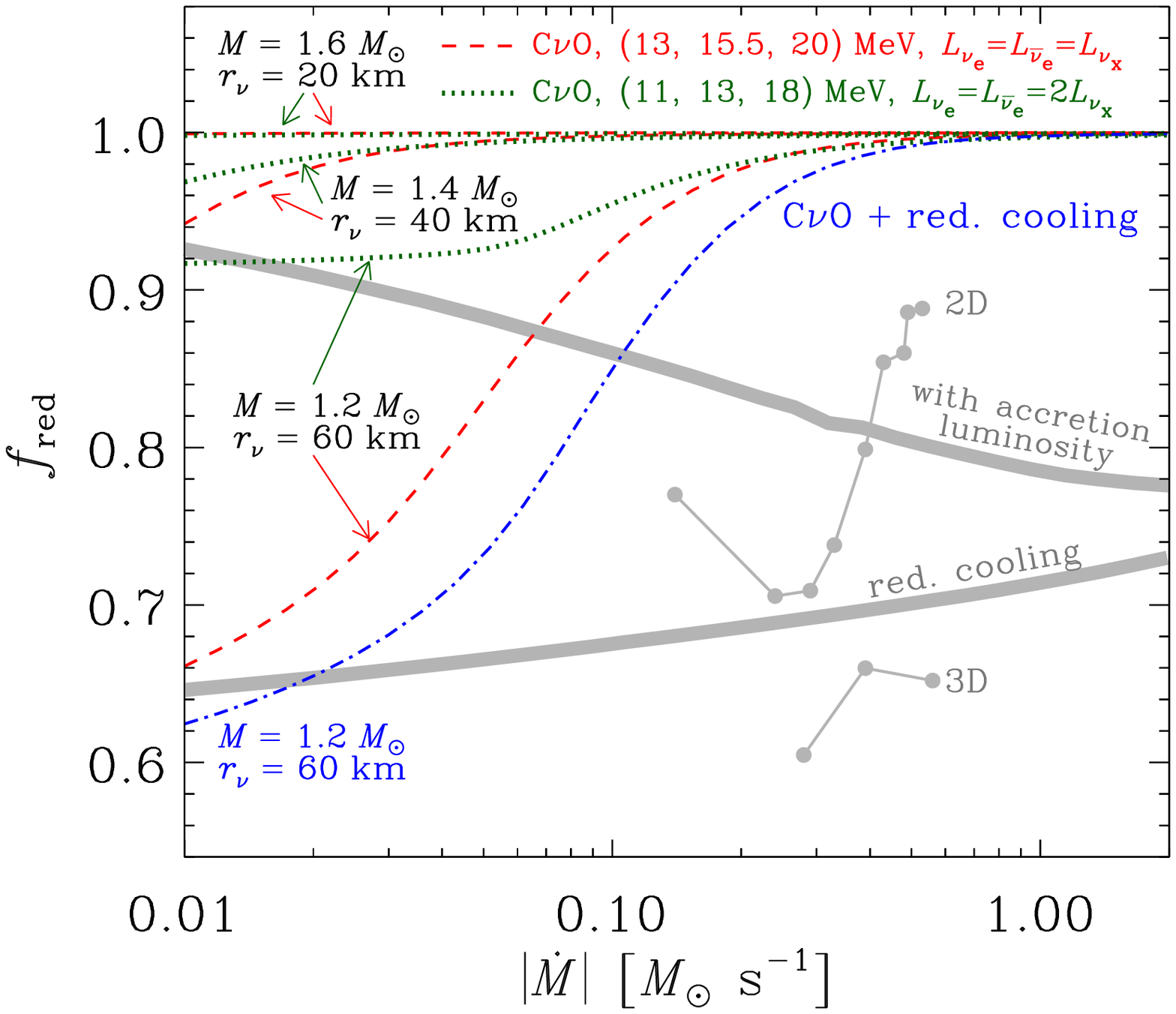}
\caption{Relative reduction of $\lcrit$ with respect to the fiducial 1D calculation ($\fred$) as a function of $\mdot$ including various physical effects. Red dashed and green dotted lines show $\fred$ for \cno\ for two sets of neutrino energies and luminosities. The legend gives $(\epsnue,\epsnuebar,\epsnux)$. Lines are labeled with $M$ and $\rnu$ of the PNS. Blue dash-dotted line shows the effect of \cno\ and reduced cooling rate by a factor of $2$ relative to a calculation with reduced cooling only. The upper thick grey line shows $\fred$ when neutrinos from cooling of the accretion flow are taken into account and the lower thick grey line illustrates $\fred$ for cooling rate reduced by a factor $2$  \citepalias{pejcha12}. The grey solid lines with points show $\fred$ for multi-dimensional effects when the dimension of the simulation is increased from 1D to 2D or from 1D to 3D \citep[from][]{nordhaus10}.}
\label{fig:fred}
\end{figure*}

How does the strength of the \cno\ effect scale with $M$ and $\rnu$? In \citetalias{pejcha12} we showed that $\rscrit$ depends predominantly on $\lcrit$ (Eq.~[\ref{eq:rscrit}]) and that the dependencies on other parameters like $M$ and $\rnu$ are much weaker. We plot values of $\rscrit$ from \citetalias{pejcha12} as dots, triangles and stars in Figure~\ref{fig:osc}, right panel, for many different $M$ and $\rnu$. We also showed in \citetalias{pejcha12} that the critical luminosity scales as a power law of $\mdot$, $M$, and $\rnu$, which we reproduce in Equation~(\ref{eq:lcrit_empir}). Thus, increasing $\mdot$ or $M$ increases $\lcrit$, which in turn decreases $\rscrit$ and increases $\rsync$ and $\rend$ (eq.~[\ref{eq:scaling}]) and the effect of \cno\ will be weaker. Similarly, lower $\rnu$ yields higher $\lcrit$ and lower $\rscrit/\rnu$. At the same time, $\rsync/\rnu$ and $\rend/\rnu$ will increase (eq.~[\ref{eq:scaling}]) and the \cno\ effect will become important at smaller $\mdot$ (smaller $\lcrit$). Therefore, the collective oscillations will be most prominent for the sets of parameters that minimize $\lcrit$: small $\mdot$, $M$, and large $\rnu$.

We plot in Figure~\ref{fig:fred} $\fred$---the ratio of $\lcrit$ including \cno\ to $\lcrit$ of a reference calculation---essentially the reduction factor of $\lcrit$ due to \cno. We note that in order to evaluate only the effect of \cno, we choose the reference calculation to have the same $M$, $\rnu$, and microphysics. Red dashed lines show $\fred$ for the neutrino parameters discussed so far. Figure~\ref{fig:fred} shows that to get a $\gtrsim 10\%$ reduction in $\lcrit$ for $\rnu=60$\,km and $M=1.2\,\msun$ due to \cno, we require $\mdot \lesssim 0.1\ \msun\ \invs$. To get the same reduction in $\lcrit$ for $M=1.4\,\msun$ and $\rnu=40$\,km, we require $\mdot \ll 0.01\ \msun\ \invs$. For $\rnu=20$\,km and $M=1.6\,\msun$, there is no noticeable reduction in $\lcrit$, because $\rscrit < \rsync$ for the whole considered range of $\mdot$. 

These results can be put into context by coupling to progenitor models. Looking at the solar-metallicity supernova progenitors of \citet{woosley02}\footnote{\url{http://www.stellarevolution.org/data.shtml}}, $\mdot=0.1\ \msun\ \invs$ is reached $\sim\! 0.65$\,s after the initiation of the collapse for an $11.2\,\msun$ progenitor, but at $\sim\! 4$\,s for a $15\,\msun$ progenitor. At these times, the accreted baryonic masses are $M=1.35$ and $2.2\,\msun$ for the $11.2$ and $15.0\,\msun$ progenitors, respectively. The lower limit of our calculations $\mdot = 0.01\ \msun\ \invs$ is reached only after $\sim\! 15$\,s for the $11.2\,\msun$ progenitor, when the PNS has almost fully cooled \citep[e.g.][]{pons99,hudepohl10,fischer10,fischer12}. From this investigation we conclude that the decrease of $\lcrit$ due to \cno\ is noticeable only for very low mass progenitors, which reach low $\mdot$ at early times, when $\rnu$ is still potentially large. This would be possible for a stiff equation of state of dense nuclear matter, which would keep $\rnu$ high\footnote{Note however, that to get an explosion $\lcrit$ has to be reached by the actual core luminosity $\lcore$, which depends on the equation of state in a more complicated way. Indeed, a softer equation of state generally leads to higher $\lcore$ at early times after bounce and may thus be favorable for explosion via the neutrino mechanism \citep[e.g.][]{marek09}.}.

The reduction of $\lcrit$ depends on the assumed neutrino energies and luminosities. So far, we have discussed the case favorable for \cno, specifically $(\epsnue^{\rm i}, \epsnuebar^{\rm i},  \epsnux^{\rm i}) = (13,15.5,20)$\,MeV and $\lcore = L_{\nuebar,{\rm core}} = L_{\nux,{\rm core}}$. Now we turn to neutrino energies and luminosities that more closely approximate the results of the recent sophisticated calculations \citep[e.g.][]{marek09,fischer10,fischer12}, namely $(\epsnue^{\rm i},\epsnuebar,\epsnux^{\rm i}) = (11,13,18)$\,MeV and $\lcore = L_{\nuebar,{\rm core}} = 2L_{\nux,{\rm core}}$, shown with green dotted lines in Figure~\ref{fig:fred}. We see that for these parameters, the \cno\ become apparent at approximately the same values of $\mdot$, but the effect is much smaller. For $M=1.2\,\msun$ and $\rnu=60$\,km the maximum possible reduction of $\lcrit$ is only about $10\%$, much smaller than $\sim 40\%$ for the equal luminosities and higher neutrino energies. Interestingly, unequal luminosities allow for a possibility of having $\fred > 1$. One case when this can happen is when $\phi_{\nux}$ is low enough, so that $\nue$ and $\nuebar$ oscillate to $\nux$, but there is very little $\nux$ to oscillate back. As a result, the luminosity in $\nue$ and $\nuebar$ decreases and \cno\ can thus be detrimental for the explosion. We do not see $\fred > 1$ for any considered parameter combination.

\subsection{Comparison to Other Effects}
\label{sec:comparison}

Now we compare the \cno\ effect to other physical effects that have been shown to decrease $\lcrit$. We have seen that the relative positions of $\rs$, $\rsync$ and $\rend$ determine the effect of \cno. It is known that multi-dimensional effects like convection and SASI consistently increase shock radii \citep[e.g.][]{bhf95,ohnishi06,iwakami08,murphy08,marek09,nordhaus10} and decrease $\lcrit$ over the corresponding 1D value \citep{murphy08,nordhaus10,suwa10,hanke11}, as illustrated by the grey lines with dots in Figure~\ref{fig:fred}. The common explanation is that multi-dimensional effects make the energy deposition of neutrinos more efficient by increasing the dwell time of the matter in the gain region \citep{murphy08,nordhaus10,takiwaki12}. In \citetalias{pejcha12}, we attempted to address this issue by adjusting the heating or cooling  within the framework of our steady-state calculations. We found that both a decrease of cooling and an increase of heating make $\lcrit$ smaller and increase $\rs$ for a fixed $\lcore$. However, $\rscrit$ increases only for the case of reduced cooling. For this reason, based on inspection of simulation results, we suggested (but did not prove) that the decrease of $\lcrit$ seen in multi-dimensional simulations is the result of less efficient neutrino cooling instead of the commonly assumed increase in heating efficiency. This decreases $\lcrit$ by about $30\%$ compared to the fiducial case, similar to the difference in $\lcrit$ observed by \citet{nordhaus10}, as evidenced by the lower thick grey line in Figure~\ref{fig:fred}. The blue dash-dotted line in Figure~\ref{fig:fred} (and in the left panel of Figure~\ref{fig:osc}) then shows $\lcrit$ with \cno\ and reduced cooling. As expected, because of lower $\lcrit$ and higher $\rs$ at fixed $\lcore$, the effect of \cno\ starts to be apparent for $\mdot \lesssim 0.5\ \msun\ \invs$ and reaches full strength for $\mdot \lesssim 0.04\ \msun\ \invs$. At low $\mdot$, the effect of \cno\ becomes comparable to that of increasing the dimension of the simulation from 1D to 3D, but only for fairly large $\rnu$ and small $M$ and for the less realistic energies and luminosities (red dashed lines).

In \citetalias{pejcha12} we investigated the effect of a simple gray neutrino radiation transport on $\lcrit$ (i.e.\ $\intd \lnu / \intd r \neq 0$). We found that including the neutrinos generated by the cooling of the accretion flow itself (the accretion luminosity) lowers $\lcrit$ by $8\%$ to $23\%$ for the mass accretion rates between $0.01$ and $2\,\msun$\,s$^{-1}$. However, the accretion luminosity was always a small fraction of $\lcrit$ and the PNS neutrino emission has to play the major role in reviving the stalled accretion shock. In Figure~\ref{fig:fred} we plot with a thick grey line $\fred$ that was obtained by including the accretion luminosity and we see that it is somewhat smaller than the maximum effect from \cno. The effect of accretion luminosity is most prominent at high $\mdot$ and has similar importance at small $\mdot$ and small $\rnu$. The effect of the accretion luminosity is comparable to going from 1D calculations to 2D in the calculations of \citet{nordhaus10}.

\subsection{Role of Multi-Angle Matter Effects}
\label{sec:multiangle}

\begin{figure}
\centering
\includegraphics[width=0.45\textwidth]{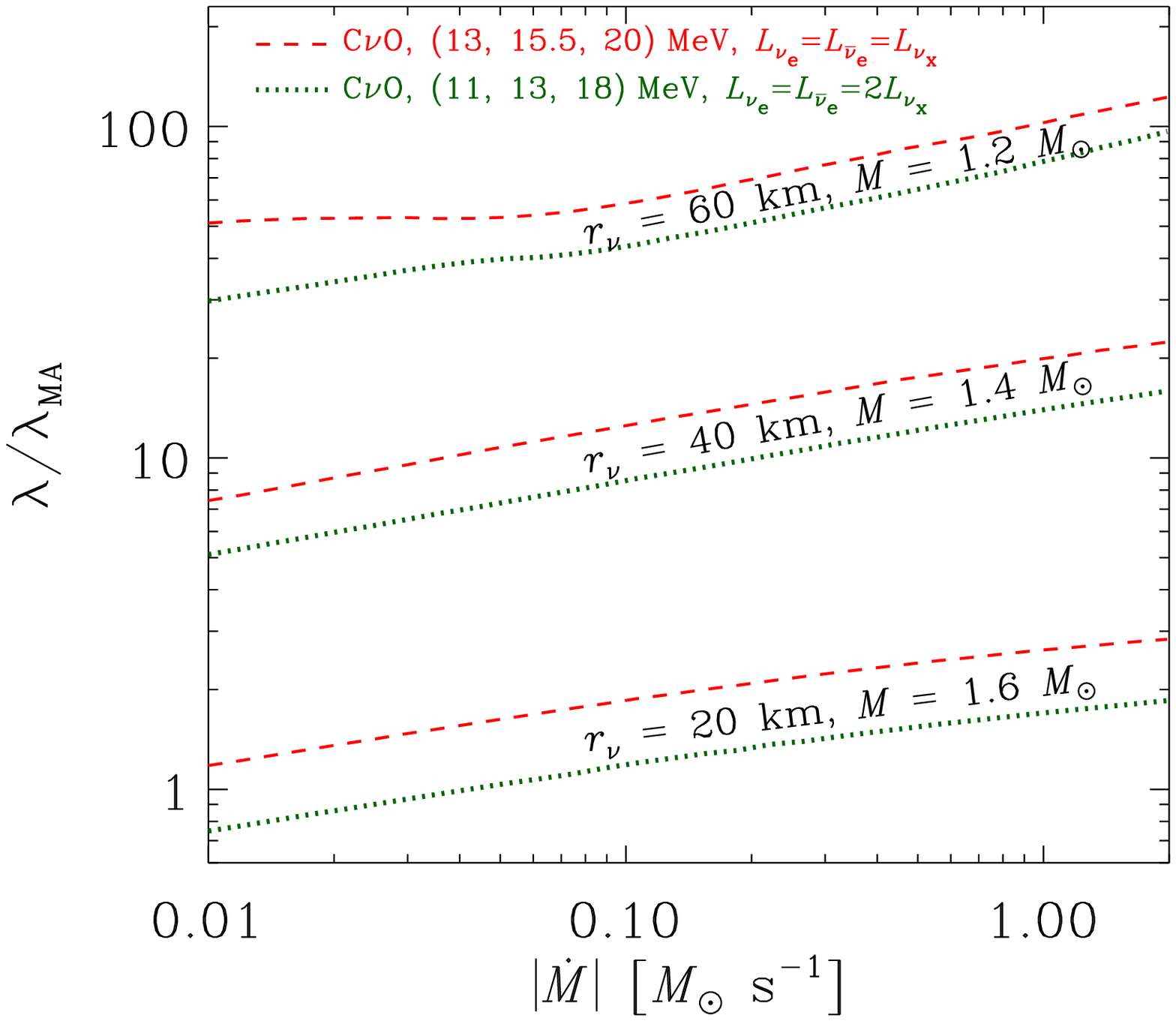}
\caption{The relative importance of the matter suppression of \cno\ parameterized by $\lambda/\lambda_{\rm MA}$ (Eqs.~[\ref{eq:lambda}]--[\ref{eq:lambda_MA}]). The lines are for the calculations with different $M$ and $\rnu$ presented with dashed red and dotted green lines in Figure~\ref{fig:fred}. The ratio $\lambda/\lambda_{\rm MA}$ was evaluated just inside of the shock and at $\lcrit$, where the ratio $\lambda/\lambda_{\rm MA}$ is smallest. If $\lambda/\lambda_{\rm MA} \gg 1$, the \cno\ are suppressed. The upward curvature of the $M=1.2\,\msun$ and $\rnu=60$\,km lines for $\mdot \lesssim 0.6\,\msun$\,s$^{-1}$ is caused by the decrease of $\lcrit$ and $\rscrit$ due to \cno\ when matter suppression effects are neglected.}
\label{fig:multiangle}
\end{figure}

Throughout this Section we have neglected the matter-suppression effects on \cno\ in order to obtain the maximum possible effect of \cno\ over a broad range of parameters. We found that \cno\ can operate only in a very restricted range in $\mdot$, $M$, and $\rnu$. At this point we evaluate the importance of matter suppression on our results. 

In Figure~\ref{fig:multiangle} we plot the ratio $\lambda/\lambda_{\rm MA}$, which estimates the relative importance of the MSW matter potential to the collective potential (Eqs.~[\ref{eq:lambda}]--[\ref{eq:lambda_MA}]). This ratio is evaluated for solutions at $\lcrit$ just inside of the shock, where it attains the smallest value. If $\lambda/\lambda_{\rm MA} \gg 1$, the collective oscillations are suppressed. We see that for large radii and small masses ($M = 1.2 \, \msun$ and $\rnu=60$\,km), the \cno\ are suppressed by the matter effects. However, this is also the parameter space where \cno\ occur for $\mdot$ attainable by low-mass progenitors together with a stiff high-density EOS (Figure~\ref{fig:fred}). Figure~\ref{fig:multiangle} shows that matter suppression effects are moderate for the other parameter combinations, but these combinations do not yield any decrease in $\lcrit$ due to \cno\ for any realistic mass accretion rates. Thus, for the region of parameter space where \cno\ are maximal, the matter suppression effects are largest, whereas where the matter suppresion is small, the effect of \cno\ is negligible.

\vspace{-0.25cm}

\section{Discussion \& Conclusions}
\label{sec:disc}

We investigate the effect of collective neutrino oscillations on the neutrino mechanism of core-collapse supernovae as parameterized by the critical neutrino luminosity $\lcrit$. We assume that neutrino energies and luminosities vary smoothly between an initial state at synchronization radius $\rsync$ and a final state end radius $\rend$, as summarized by \citet{dasgupta12}. The final states are dictated by the neutrino number conservation. Without matter-suppression, we found that collective oscillations affect $\lcrit$ if $\rsync < \rscrit$, where $\rs^{\rm crit}$ is the shock radius at $\lcrit$, and the full magnitude of the effect occurs if $\rend < \rscrit$.

The reduction of $\lcrit$ depends on the assumed energy difference between $\nue$, $\nuebar$ and $\nux$, and on the individual luminosities. We find that neutrino energies of \citet{thompson03} and equal luminosities in each flavor are favorable for \cno, giving reduction of $\lcrit$ by a factor of $\sim\! 1.5$ ($\fred = 0.65$, Fig.~\ref{fig:fred}). The more recent calculations of \citet{marek09} and \citet{fischer10,fischer12} predict slightly lower energies and approximately half the $\nux$ luminosity compared to $\nue$ and $\nuebar$. For these parameter values we find that the $\lcrit$ reduction reaches only about $10\%$, but the parameter space in terms of $\mdot$, $M$, and $\rnu$ is essentially the same. We do not find any increase of $\lcrit$ for the range of parameters considered. Conversely, if the energy $\epsnux^{\rm i}$ was increased to $25$\,MeV or $35$\,MeV while keeping $\epsnue^{\rm i}=13$\,MeV, $\epsnuebar^{\rm i}=15.5$\,MeV, and luminosities of all flavors equal, $\lcrit$ at low $\mdot$ would be reduced by a factor of $2$ and $3.5$ ($\fred \simeq 0.5$ and $0.3$), respectively.

We find that \cno\ can be important only for low $\mdot$, small $M$, and large $\rnu$ (see Fig.~\ref{fig:osc}, left panel, red lines). This is best achieved in the lowest mass progenitors, in which $\mdot$ decreases very rapidly due to their steep density structure. However, for times $\lesssim 0.65$\,s after the collapse is initiated, $\mdot$ is likely still too high to cause a \cno-driven explosion in an $11.2\,\msun$ progenitor \citep[see also][]{chakraborty11a,chakraborty11b,dasgupta12}. At late times, $\mdot$ decreases rapidly, but the effect of \cno\ on $\lcrit$ is likely offset by the simultaneous drop of $\rnu$ as the PNS cools and the concomitant increase in $M$. Figure~\ref{fig:fred} shows that for smaller $\rnu$ and larger $M$, \cno\ does not have any effect. Thus, only if the decrease in $\rnu$ is slow enough, perhaps due to a stiff equation of state, $\mdot$ might drop enough so that the decrease in $\lcrit$ due to \cno\ can potentially influence the system before black hole formation or explosion via the ordinary neutrino mechanism. However, as is indicated in Figure~\ref{fig:multiangle}, these parameter combination are the ones most affected by matter suppression and \cno\ are unlikely.

Finally, we note that even in our implementation of \cno\ the parameter space where \cno\ could operate is small and reduced even more due to matter suppression effects. Realistically, the potential of \cno\ flavor conversion may be reduced even more. In particular, refinements in the treatment of the \cno\ typically make the effect of \cno\ still smaller, because the physical scale of conversion is moved to larger radii due to various multi-angle effects \citep{esteban08,chakraborty11a,chakraborty11b,dasgupta12,sarikas11}. We find that the multi-angle effects are important especially for small $M$ and large $\rnu$ (Figure~\ref{fig:multiangle}), where \cno\ have the greatest potential of influencing the supernova explosion. If the physical scale of conversion is pushed to a radius larger than $\rscrit$ for a given $\mdot$, $M$, $\rnu$, the \cno\ will not have any effect on $\lcrit$. We thus conclude that \cno\ are unlikely to play a role in neutrino-driven explosions by evaluating these effects at the critical luminosity.

\vspace{-0.25cm}

\section*{Acknowledgments}
This work is supported in part by an Alfred P. Sloan Foundation Fellowship and by NSF grant AST-0908816. We thank John Beacom, Evan O'Connor, Christian Ott, Hans-Thomas Janka, and Alessandro Mirizzi. We also thank the anonymous referee for comments and suggestions that helped to improve the paper.


\vspace{-0.25cm}


\begin{thebibliography}{}
\footnotesize
\bibliographystyle{mn2e} 
\bibitem[Akiyama et al.(2003)]{akiyama03} Akiyama, S., Wheeler, J.~C., Meier, D.~L., \& Lichtenstadt, I.\ 2003, \apj, 584, 954 
\bibitem[Bethe \& Wilson(1985)]{bethewilson85} Bethe, H.~A., \& Wilson, J.~R.\ 1985, \apj, 295, 14 
\bibitem[Bruenn(1989a)]{bruenn89a} Bruenn, S.~W.\ 1989a, \apj, 340, 955 
\bibitem[Bruenn(1989b)]{bruenn89b} Bruenn, S.~W.\ 1989b, \apj, 341, 385 
\bibitem[Bruenn \& Dineva(1996)]{bruenn96} Bruenn, S.~W., \& Dineva, T.\ 1996, \apjl, 458, L71 
\bibitem[Bruenn et al.(2001)]{bruenn01} Bruenn, S.~W., De Nisco, K.~R., \& Mezzacappa, A.\ 2001, \apj, 560, 326 
\bibitem[Buras et al.(2006)]{buras06a} Buras, R., Rampp, M., Janka, H.-T., \& Kifonidis, K.\ 2006a, \aap, 447, 1049 
\bibitem[Burrows \& Lattimer(1985)]{burrows85} Burrows, A., \& Lattimer, J.~M.\ 1985, \apjl, 299, L19 
\bibitem[Burrows(1987)]{burrows87} Burrows, A.\ 1987, \apjl, 318, L57 
\bibitem[Burrows \& Goshy(1993)]{bg93} Burrows, A., \& Goshy, J.\ 1993, \apjl, 416, L75
\bibitem[Burrows et al.(1995)]{bhf95} Burrows, A., Hayes, J., \& Fryxell, B.~A.\ 1995, \apj, 450, 830 
\bibitem[Burrows et al.(2006)]{burrows06} Burrows, A., Livne, E., Dessart, L., Ott, C.~D., \& Murphy, J.\ 2006, \apj, 640, 878 
\bibitem[Chakraborty et al.(2011a)]{chakraborty11a}  Chakraborty, S., Fischer, T., Mirizzi, A., Saviano, N., \& Tom{\`a}s, R.\ 2011a, Physical Review Letters, 107, 151101  
\bibitem[Chakraborty et al.(2011b)]{chakraborty11b} Chakraborty, S., Fischer, T., Mirizzi, A., Saviano, N., \& Tom{\`a}s, R.\ 2011b, \prd, 84, 025002  
\bibitem[Colgate \& White(1966)]{colgate66} Colgate, S.~A., \& White, R.~H.\ 1966, \apj, 143, 626 
\bibitem[Dasgupta et al.(2009)]{dasgupta09} Dasgupta, B., Dighe, A., Raffelt, G.~G., \& Smirnov, A.~Y.\ 2009, Physical Review Letters, 103, 051105 
\bibitem[Dasgupta et al.(2012)]{dasgupta12} Dasgupta, B., O'Connor, E.~P., \& Ott, C.~D.\ 2012, \prd, 85, 065008
\bibitem[Dessart et al.(2008)]{dessart08} Dessart, L., Burrows, A., Livne, E., \& Ott, C.~D.\ 2008, \apjl, 673, L43  
\bibitem[Duan et al.(2006)]{duan06} Duan, H., Fuller, G.~M., Carlson, J., \& Qian, Y.-Z.\ 2006, \prd, 74, 105014 
\bibitem[Duan et al.(2010)]{duan10} Duan, H., Fuller, G.~M., \& Qian, Y.-Z.\ 2010, Annual Review of Nuclear and Particle Science, 60, 569 
\bibitem[Esteban-Pretel et al.(2007)]{esteban07} Esteban-Pretel, A., Pastor, S., Tom{\`a}s, R., Raffelt, G.~G., \& Sigl, G.\ 2007, \prd, 76, 125018 
\bibitem[Esteban-Pretel et al.(2008)]{esteban08} Esteban-Pretel, A., Mirizzi, A., Pastor, S., Tom{\`a}s, R., Raffelt, G.~G., Serpico, P.~D., \& Sigl, G.\ 2008, \prd, 78, 085012 
\bibitem[Fern{\'a}ndez(2012)]{fernandez12} Fern{\'a}ndez, R.\ 2012, \apj, 749, 142 
\bibitem[Fischer et al.(2010)]{fischer10} Fischer, T., Whitehouse, S.~C., Mezzacappa, A., Thielemann, F.-K., \& Liebend{\"o}rfer, M.\ 2010, \aap, 517, A80 
\bibitem[Fischer et al.(2012)]{fischer12} Fischer, T., Mart{\'{\i}}nez-Pinedo, G., Hempel, M., \& Liebend{\"o}rfer, M.\ 2012, \prd, 85, 083003 
\bibitem[Fryer \& Warren(2004)]{fryer04} Fryer, C.~L., \& Warren, M.~S.\ 2004, \apj, 601, 391 
\bibitem[Hanke et al.(2011)]{hanke11} Hanke, F., Marek, A., Mueller, B., \& Janka, H.-T.\ 2011, arXiv:1108.4355 
\bibitem[Hannestad et al.(2006)]{hannestad06} Hannestad, S., Raffelt, G.~G., Sigl, G., \& Wong, Y.~Y.~Y.\ 2006, \prd, 74, 105010 
\bibitem[Herant et al.(1994)]{herant94} Herant, M., Benz, W., Hix, W.~R., Fryer, C.~L., \& Colgate, S.~A.\ 1994, \apj, 435, 339 
\bibitem[H{\"u}depohl et al.(2010)]{hudepohl10} H{\"u}depohl, L., M{\"u}ller, B., Janka, H.-T., Marek, A., \& Raffelt, G.~G.\ 2010, Physical Review Letters, 104, 251101 
\bibitem[Iwakami et al.(2008)]{iwakami08} Iwakami, W., Kotake, K., Ohnishi, N., Yamada, S., \& Sawada, K.\ 2008, \apj, 678, 1207 
\bibitem[Janka \& M\"uller(1996)]{janka96} Janka, H.-T., \& M\"uller, E.\ 1996, \aap, 306, 167 
\bibitem[Janka et al.(2005)]{janka05} Janka, H.-T., Buras, R., Kitaura Joyanes, F.~S., Marek, A., Rampp, M., \& Scheck, L.\ 2005, Nuclear Physics A, 758, 19 
\bibitem[Janka et al.(2008)]{janka08} Janka, H.-T., M{\"u}ller, B., Kitaura, F.~S., \& Buras, R.\ 2008, \aap, 485, 199 
\bibitem[Keil et al.(1996)]{keil96} Keil, W., Janka, H.-T., \& Mueller, E.\ 1996, \apjl, 473, L111 
\bibitem[Kitaura et al.(2006)]{kitaura06} Kitaura, F.~S., Janka, H.-T., \& Hillebrandt, W.\ 2006, \aap, 450, 345
\bibitem[LeBlanc \& Wilson(1970)]{leblanc70} LeBlanc, J.~M., \& Wilson, J.~R.\ 1970, \apj, 161, 541  
\bibitem[Liebend{\"o}rfer et al.(2001)]{liebendorfer01} Liebend{\"o}rfer, M., Mezzacappa, A., Thielemann, F.-K., Messer, O.~E., Hix, W.~R., \& Bruenn, S.~W.\ 2001, \prd, 63, 103004 
\bibitem[Marek \& Janka(2009)]{marek09} Marek, A., \& Janka, H.-T.\ 2009, \apj, 694, 664 
\bibitem[Mezzacappa et al.(2001)]{mezzacappa01} Mezzacappa, A., Liebend{\"o}rfer, M., Messer, O.~E., Hix, W.~R., Thielemann, F.-K., \& Bruenn, S.~W.\ 2001, Physical Review Letters, 86, 1935
\bibitem[M\"uller et al.(2012)]{muller12} M\"uller, B., Janka, H.-T., \& Marek, A.\ 2012, arXiv:1202.0815  
\bibitem[Pons et al.(1999)]{pons99} Pons, J.~A., Reddy, S., Prakash, M., Lattimer, J.~M., \& Miralles, J.~A.\ 1999, \apj, 513, 780 
\bibitem[Murphy \& Burrows(2008)]{murphy08} Murphy, J.~W., \& Burrows, A.\ 2008, \apj, 688, 1159 
\bibitem[Nordhaus et al.(2010)]{nordhaus10} Nordhaus, J., Burrows, A., Almgren, A., \& Bell, J.\ 2010, \apj, 720, 694 
\bibitem[Ohnishi et al.(2006)]{ohnishi06} Ohnishi, N., Kotake, K., \& Yamada, S.\ 2006, \apj, 641, 1018 
\bibitem[Pantaleone(1992)]{pantaleone92} Pantaleone, J.\ 1992, Physics Letters B, 287, 128 
\bibitem[Pejcha \& Thompson(2012)]{pejcha12} Pejcha, O., \& Thompson, T.~A.\ 2012, \apj, 746, 106 
\bibitem[Qian \& Woosley(1996)]{qian96} Qian, Y.-Z., \& Woosley, S.~E.\ 1996, \apj, 471, 331 
\bibitem[Rampp \& Janka(2000)]{rampp00} Rampp, M., \& Janka, H.-T.\ 2000, \apjl, 539, L33 
\bibitem[Sarikas et al.(2011)]{sarikas11} Sarikas, S., Raffelt, G.~G., H{\"u}depohl, L., \& Janka, H.-T.\ 2012, Physical Review Letters, 108, 061101 
\bibitem[Scheck et al.(2006)]{schecketal06} Scheck, L., Kifonidis, K., Janka, H.-T., M\"{u}ller, E.\ 2006, \aap, 457, 963 
\bibitem[Suwa et al.(2010)]{suwa10} Suwa, Y., Kotake, K., Takiwaki, T., et al.\ 2010, \pasj, 62, L49 
\bibitem[Suwa et al.(2011)]{suwa11} Suwa, Y., Kotake, K., Takiwaki, T., Liebend{\"o}rfer, M., \& Sato, K.\ 2011, \apj, 738, 165
\bibitem[Symbalisty(1984)]{symbalisty84} Symbalisty, E.~M.~D.\ 1984, \apj, 285, 729   
\bibitem[Takiwaki et al.(2012)]{takiwaki12} Takiwaki, T., Kotake, K., \& Suwa, Y.\ 2012, \apj, 749, 98  
\bibitem[Thompson et al.(2003)]{thompson03} Thompson, T.~A., Burrows, A., \& Pinto, P.~A.\ 2003, \apj, 592, 434
\bibitem[Thompson et al.(2005)]{thompson05} Thompson, T.~A., Quataert, E., \& Burrows, A.\ 2005, \apj, 620, 861  
\bibitem[Wilson \& Mayle(1988)]{wilson88} Wilson, J.~R., \& Mayle, R.~W.\ 1988, \physrep, 163, 63 
\bibitem[Woosley et al.(2002)]{woosley02} Woosley, S.~E., Heger, A., \& Weaver, T.~A.\ 2002, Reviews of Modern Physics, 74, 1015 
\bibitem[Yamasaki \& Yamada(2005)]{yamasaki05} Yamasaki, T., \& Yamada, S.\ 2005, \apj, 623, 1000 
\bibitem[Yamasaki \& Yamada(2006)]{yamasaki06} Yamasaki, T., \& Yamada, S.\ 2006, \apj, 650, 291 
\bibitem[Y{\"u}ksel \& Beacom(2007)]{yuksel07} Y{\"u}ksel, H., \& Beacom, J.~F.\ 2007, \prd, 76, 083007 
\end{thebibliography}
\end{document}